\documentclass[10pt,conference]{IEEEtran}

\usepackage[utf8]{inputenc}

\usepackage{fullpage}

\usepackage{amssymb}
\usepackage{boxedminipage}

\usepackage{listings}

\usepackage{ifpdf}

\usepackage{graphicx}
\usepackage{paralist}
\usepackage{placeins}
\usepackage{url}

\begin{document}
	
% \begin{frontmatter}

\title{Finding a boundary between valid and invalid regions of the input space}

\author{ 
\IEEEauthorblockN{Bogdan Marculescu\IEEEauthorrefmark{1}\\
Robert Feldt\IEEEauthorrefmark{1}\IEEEauthorrefmark{2}}

\IEEEauthorblockA{\IEEEauthorrefmark{1}Blekinge Institute of Technology\\
School of Computing\\
Karlskrona, Sweden}

\IEEEauthorblockA{\IEEEauthorrefmark{2}Chalmers and the University of Gothenburg\\
Dept.\ of Computer Science and Engineering\\
Gothenburg, Sweden}
}

\date{}

\ifpdf
\DeclareGraphicsExtensions{.pdf, .jpg, .tif}
\else
\DeclareGraphicsExtensions{.eps, .jpg}
\fi

\maketitle

\begin{abstract}
	
%\textbf{Context}: 

In the context of robustness testing, the boundary between the valid and invalid regions of the input space can be an interesting source of erroneous inputs. Knowing where a specific software under test (SUT) has a boundary is also essential for validation in relation to requirements. However, finding where a SUT actually implements the boundary is a non-trivial problem that has not gotten much attention.

%\textbf{Objective}: 
This paper proposes a method of finding the boundary between the valid and invalid regions of the input space, by developing pairs of test sets that describe that boundary in detail. 

%\textbf{Method}: 
The proposed method consists of two steps. First, test data generators, directed by a search algorithm to maximise distance to known, valid test cases, generate valid test cases that are closer to the boundary. Second, these valid test cases undergo mutations to try to push them over the boundary and into the invalid part of the input space. This results in a pair of test sets, one consisting of test cases on the valid side of the boundary and a matched set on the outer side, with only a small distance between the two sets. The method is evaluated on a number of examples from the standard library of a modern programming language.

We propose a method of determining the boundary between valid and invalid regions of the input space, and apply it on a SUT that has a non-contiguous valid region of the input space. From the small distance between the developed pairs of test sets, and the fact that one test set contains valid test cases and the other invalid test cases, we conclude that the pair of test sets described the boundary between the valid and invalid regions of that input space.

Differences of behaviour can be observed between different distances and different sets of mutation operators, but all show that the method is able to identify the boundary between the valid and invalid regions of the input space. This is an important step towards more automated robustness testing.
	
\end{abstract}

\section{Introduction} % (fold)
\label{sec:introduction}
	
	In their introduction to software testing, Ammann and Offutt~\cite{ammann2016introduction} describe the problem of input domain representation. Different stages of software development have different notions of what the input domain is, and where the boundary between the valid and invalid input spaces lies. This region of the input space is identified as a rich source of software errors. 
	
	In this paper we will focus on robustness, as a non-functional quality criterion. We will use the definition of robustness proposed by Avi\v{z}enis~\cite{avizienis2001fundamental}: ``dependability with respect to erroneous inputs'', a deeper discussion of robustness testing can be found in \cite{shahrokni2013systematic}. In line with Ammann and Offutt~\cite{ammann2016introduction}, we argue that the boundary between the valid and invalid input spaces is a rich source of inputs that could be considered valid at one stage of development, e.g.\ specification, but invalid in the implementation and thus could be considered erroneous inputs. This is related to previous work on robustness and fuzz testing, such as Ballista~\cite{kropp1998automated} and JCrasher~\cite{csallner2004jcrasher}, and also to methods for creating test diversity~\cite{Nikolik2006,Feldt2008,Feldt2016}. However, to the best of our knowledge, few methods exist to actively seek this boundary out, rather than using ideas based on random testing or seeking test diversity, in general.
	
	For a software system, the set of test cases typically focuses around previously known issues, or inputs that developers assume could be problematic. Such typical test cases usually only cover a small part of the input space. For robustness testing, typical test cases offer a very good starting point, but are not really sufficient to investigate the behaviour of a system at the limit of its accepted valid input space.
	
	To enable a better evaluation of system robustness, we have to look at both atypical, but valid, test cases and invalid test cases. However, no guarantee can be made that the valid region of the input space is contiguous. This could mean that methods that rely on mutating existing test cases could become stuck in a local optimum and be unable to reach other regions of valid inputs. This would leave those additional regions, and the boundaries surrounding them, unexplored. 
	
\begin{figure}
    	\centering
    		\includegraphics[scale=0.5]{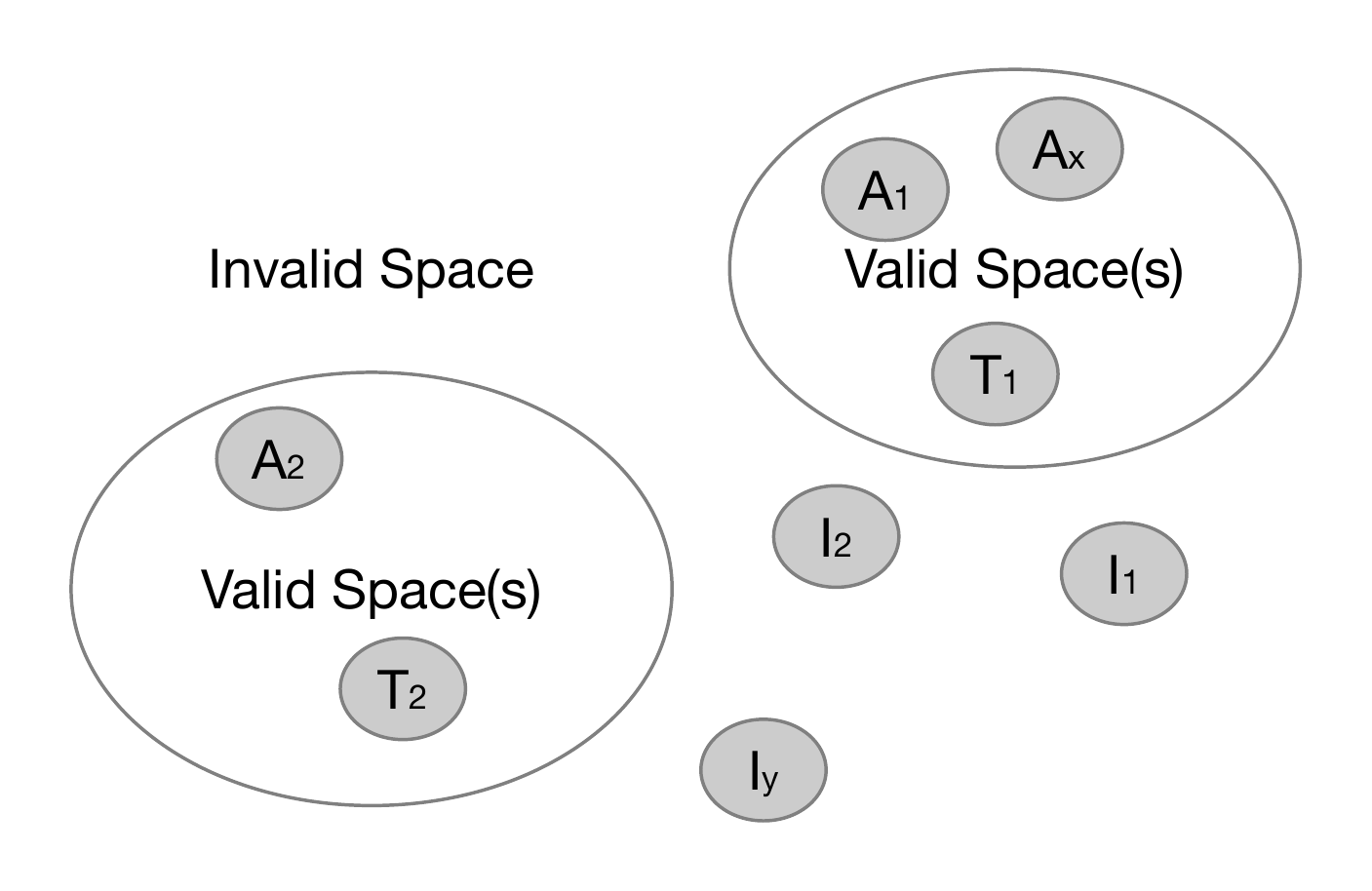}
    	\caption{Overview of the test sets. Test sets $T_1$ and $T_2$ show the typical test sets. Test sets $A_1, A_2, A_x$ comprise atypical (but still valid) tests. Test sets $I_1, I_2, I_y$ represent test cases that have invalid inputs.}
    	\label{fig:testspace}
    \end{figure}

	Figure~\ref{fig:testspace} shows a hypothetical input space containing two valid input regions that are not contiguous. Test cases $T_1$ and $T_2$ represent typical test cases that are usually included in existing test suites. Test cases $I_1$, $I_2$, and $I_y$ are examples of test cases with invalid inputs. Test cases $A_1$, $A_2$, and $A_x$ are examples of test cases that have valid inputs, but are considered atypical, i.e.\ not usually present in the test suites, not likely to be generated by existing methods, but interesting from the perspective of robustness testing. 
	
	Random testing offers a way of covering the input space, and some of the generated test cases are likely to be close to the valid-invalid boundary. However, it is difficult, for a general case, to determine which of the generated inputs are likely to prove interesting from a robustness testing perspective.

    The initial results of Poulding and Feldt~\cite{poulding2017} proposed the use of test data generators, mutations and search to explore the input space for robustness testing. Once the valid input space has been explored, mutation operators can be used to create potentially invalid inputs from the valid set. The exact boundary between the valid and invalid sets can be assumed to be between the initial (valid) test input and the mutated (invalid) one. 
    
    Here, we continue this line of work and propose a direct method of finding the boundary between the valid and invalid regions of the input space. The method results in a pair of test sets, containing test cases on either side of the valid-invalid boundary.
	
% section introduction (end)

\section{Proposed approach} % (fold)
\label{sec:proposed}

As stated previously, the goal of the method proposed in this paper is to find test cases that are close to the boundary between the valid and invalid regions of the test space. A good description of the boundary would be provided by a pair of test sets, one valid and one invalid, with a relatively small distance between them.

%Random generation of test cases has a reasonable chance of producing test cases that exist close to the boundary. However, for a generic problem it is difficult to determine which of those test cases are closest to the boundary. To find the boundary, we would be required to create test cases that are consistently close to that boundary. 

For the purpose of this work, we consider valid those inputs that the current implementation of the SUT can process correctly, i.e.\ without crashing or throwing exceptions. This definition does not depend on the system then exhibiting ``correct'' behaviour, but rather on the system being able to continue operating. 

The method we propose consists of two steps. The first step is to use automated data generation to explore the valid region of the space, and in particular to find test cases that are as distant as possible from each other. This will create a set of valid test cases that are likely to be close to the boundary. 

The second step is to mutate the valid test cases, using property switching search, to cross the valid-invalid boundary repeatedly. Small mutations would allow the resulting test cases to cross the valid-invalid boundary repeatedly, this would lead to paired valid-invalid test sets that would be close to each other.

\subsection{Step 1: Finding the boundary} % (fold)
\label{sub:step1}

The first step of the method is to use G\"{o}delTest~\footnote{https://github.com/simonpoulding/DataGenerators.jl} to develop an automated test data generator for the given SUT. G\"{o}delTest is a framework for developing automated test data generators. It has been implemented for the Julia~\footnote{https://julialang.org/} programming language. G\"{o}delTest allows the development of automated test data generators that generate correctly formed complex inputs in a non-deterministic way~\cite{poulding2015xml}. 

We assume that this generator is fault-free, generates only valid data for the current system under test, and has the potential to generate all possible valid data, given enough time and resources. 

In practical use, the generators would be based on the input space as defined by the conceptual image of the SUT: software requirements specifications, customer input, domain knowledge, etc. The inputs developed by the generator are validated by running the implemented code in Step 2, as described in Section~\ref{sub:step2}.

\begin{figure}
	\centering
		\includegraphics[scale=0.3]{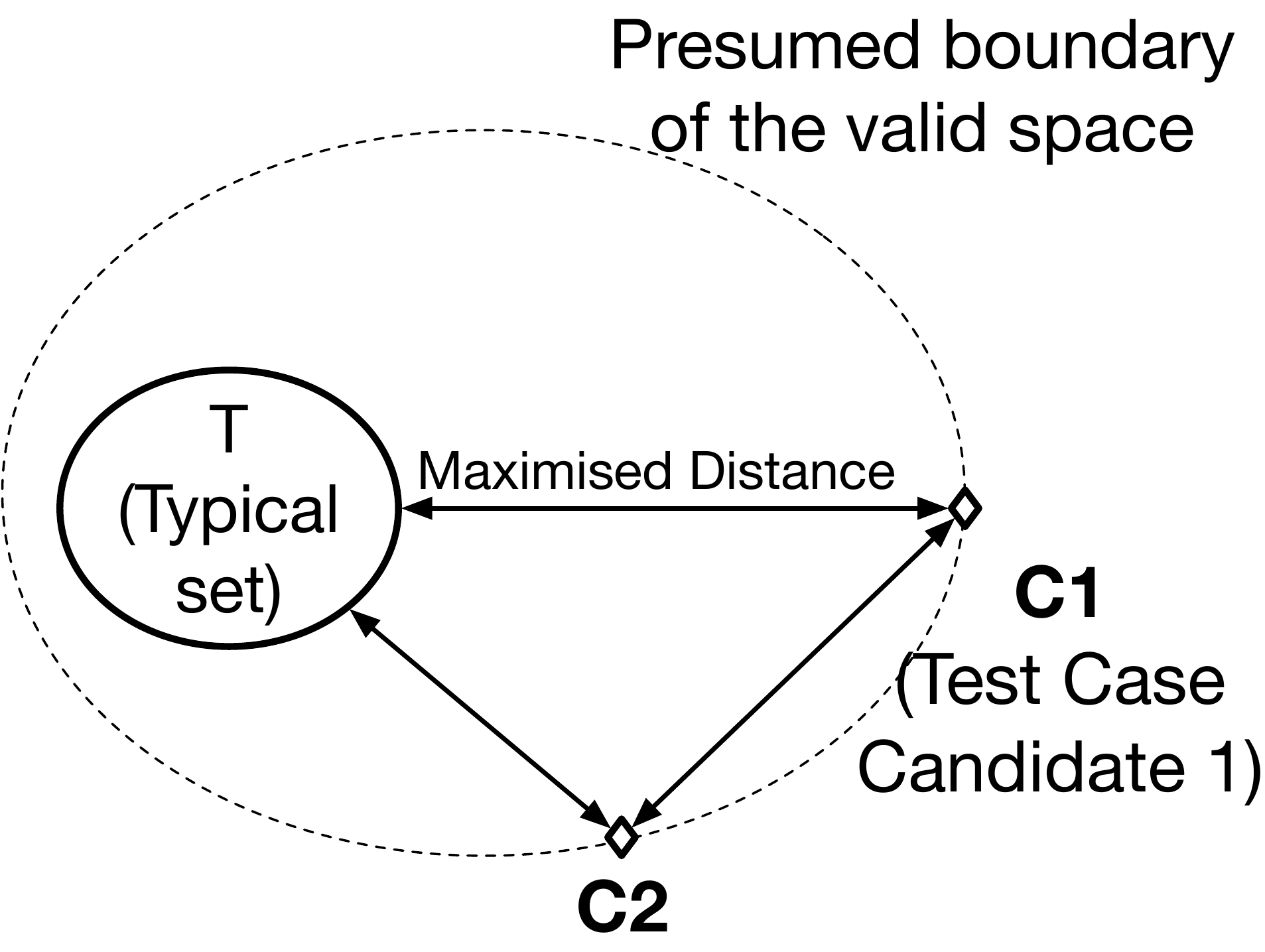}
	\caption{Step 1: Finding the inner boundary of the valid set, as defined by the generator.}
	\label{fig:step1}
\end{figure}

More specifically, the first step is to use the generator to determine what is the inner boundary of the valid set of inputs, as can be seen in Figure~\ref{fig:step1}. G\"{o}delTest offers the option of optimising the generation process, using a Nested Monte-Carlo Search (NMCS). To drive the search, a fitness function is defined, based on the distance between the current candidate $C_i$ and the reunion of the typical test set and the previously accepted candidates $Ct = {C_1, C_2,..., C_{i-1}}$. We will refer to this reunion as the $Tset$. This fitness function is to be maximised.

\begin{equation}
	\mathrm{FF}_{i} = min{dist(C_i, Tset)}
	\label{eq:iff}
\end{equation}

In other words, the search will find the candidate for which the minimum distance between itself and the set of candidates already found is the highest. This will push the search towards previously unexplored areas of the input set.

Once a candidate $C_i$ is selected, as a result of the NMCS and proposed, it is added to the $Tset$, and the process resumes. The end result of Step 1 is a set of candidate test cases that are assumed to be:
\begin{itemize}
    \item Valid inputs, but close to the boundary between the valid and invalid spaces.
    \item Spread out across the boundary between the valid and invalid spaces.
\end{itemize}

Step 1 relies only on the definition of the input space, and does not require working software.

The result of Step 1 is the set of test cases $Tset$. The $Tset$ contains a number of valid test cases that are automatically generated to be as close as possible to the valid-invalid boundary. Since the generator is assumed to only propose valid test cases, the $Tset$ on the valid side of the boundary. Each of the individual elements of the $Tset$ forms a starting point for the next step. 

% subsection step1 (end)

\subsection{Step 2: Crossing into the invalid set} % fold
\label{sub:step2}

The second step starts from the individual test cases in the $Tset$, and relies on mutation to cross the valid-invalid boundary. The result of the second step consists of close, paired, sets of test cases on either side of the boundary. 

For each candidate in the set $Tset$ of accepted test candidates close to the boundary, we conduct a property switching search. 

The property switching search starts from a valid candidate, and an automated means of checking if the candidate exhibits a desired property. The candidate is mutated until the property changes value. The new candidate undergoes the same process until the property changes value again. The result of the property switching search is the creation of two sets of test cases, one that has a desired property and one that does not, that have all resulted from mutating the same starting candidate.

In our case, the desired property is validity. We then conduct step-wise mutations, until the resulting mutant becomes invalid. The invalid test case is added to the set of invalid cases, and then further modified until its property switches back to valid, and so on. Each test case is evaluated against the implemented software to check its validity.

%In a general sense, we define a search type, property switching search, that searches for mutants for which some property is alternately true and false. The results of applying this search are two sets of test case candidates, obtained from each other using step-wise mutations. One of these sets will consist of test cases that have the property, and the other of test cases that do not. In our implementation, the property being investigated is validity. 

\begin{figure}
	\centering
		\includegraphics[scale=0.23]{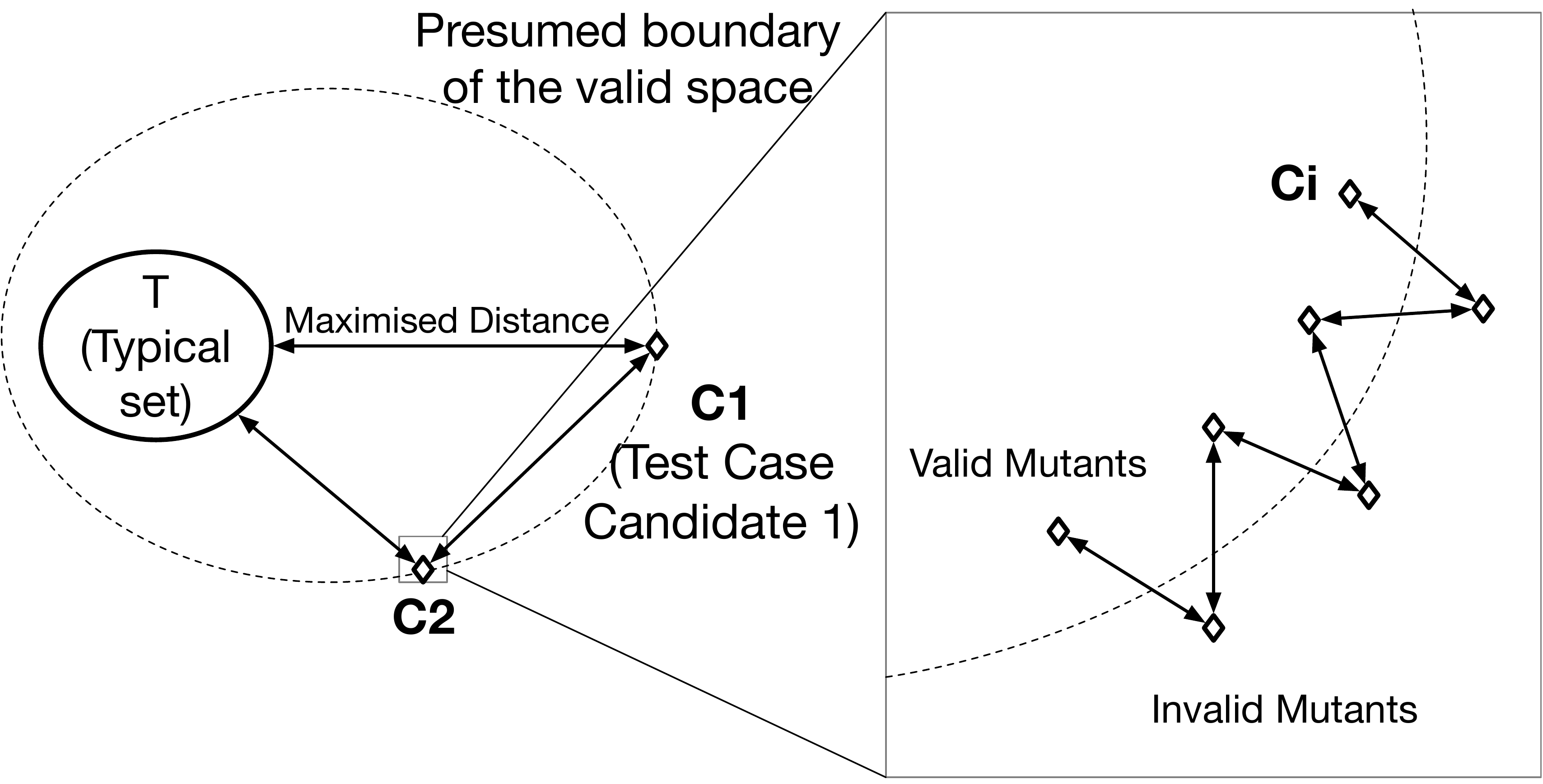}
	\caption{Step 2: Oscillating between valid and invalid candidates that can be obtained through one basic mutation step. The boundary between the valid and invalid spaces, in this region of the input space, can be assumed to be between the two sets of candidates.}
	\label{fig:step2}
\end{figure}

The second step results in two sets of test cases, one valid and one invalid, that are close to each other. Thus, we can assume that for the regions of the input space being investigated, the boundary between the valid and invalid spaces lies between the two sets. 

% subsection step2 (end)

% section proposed (end)

\section{Validation} % (fold)
\label{sec:validation}

The initial effort at validation will focus on automated generation of calendar dates, for use with the DateParser package of the Julia language. The DateParser package we will assess takes as input a string and attempts to parse it into a Date object. An input string is valid if it can successfully be parsed into a Date object. The DateParser throws an error when the input string cannot be parsed, i.e.\ when the input is invalid. Date generation has the benefit of being a comparatively simple type of input, but one that is structured and easily understood. 

Moreover, date generation is interesting because the valid-invalid boundary describes non-contiguous regions of valid space, examples of which can be seen in Figure~\ref{fig:testspace}. Invalid sets of dates, ``close'' to the boundary can also be generated based on our understanding of the domain, in this case of calendar dates.

\subsection{A word on distances} % (fold)
\label{sub:distances}

Both steps of the proposed approach rely on measurements of distance. The first step optimises generated data by maximising the distance between the new datum and the existing data. The second looks at finding paired valid-invalid sets with small distances between them. 

For the purposes of this study, we define three types of distance:
\begin{itemize}
    \item \textbf{Domain agnostic:} This is a distance that is generalizable and applicable regardless of domain. Normalised compression distance (NCD) is such a distance metric. NCD is an approximation of the Kolmogorov distance that seeks to capture the difference in information, regardless of the information's encoding. NCD is extensively used in Information Theory and has been shown to have a number of useful properties~\cite{ncd_entity_ident, ncd_noise}.
    
    \item \textbf{Domain agnostic, but encoding specific:} This describes distances that are specific to particular data representations, but not domains. In our case, we look at SUTs that use string inputs. As a result, a metric that evaluates the distance between the string inputs would be useful, regardless of the semantics of the data being encoded. Levenshtein distance~\cite{levenshtein1966binary} describes the similarity of strings. While this distance is domain agnostic (many different domains can store data in text form) it is specific to the string encoding.
    
    \item \textbf{Domain specific:} This type of distance metric refers to distances that rely on knowledge of the domain, of the data being encoded, and of the potential use of that data. This is problematic as that knowledge may be difficult to validate, but could potentially better describe the data in question. In our case we define two domain specific distances. Day Distance attempts to describe the distance (in days) between two dates, based on their string representation. Most Significant Int Distance (MSID) describes the difference between the most significant integer value in each string representation. These are somewhat artificial examples of a domain specific distances. The goal is to show that such domain specific distances (that might be required in particular settings) can be used to drive G\"{o}delTest.
\end{itemize}

Domain agnostic distance metrics, especially NCD, are likely to be computationally expensive. This could be a problem if pairwise comparisons between large data sets are required. However, such distance metrics are readily generalizable and applicable on other SUTs and even other domains. Thus, such metrics are likely to be computationally expensive, but require no additional resources for development or validation. 

Domain agnostic, but representation specific distances, like our example of Levensthein distance, can be generalised to SUTs that use the same type of representation. Since strings are a common representation of input data, we can assume that using such distance metrics would lead to a generalizable system, albeit one with more severe limitations. 

For both these types of metrics, implementations and toolkit support already exist. These distances are likely to have a lower barrier to entry since they are already studied, their applicability and generalizability more broadly investigated, and their implementations more likely to be fault-free and optimised.

Domain specific distances are likely to be useful for particular problems. Understanding peculiarities of the domain, or even of individual SUTs, can lead to better optimised distances that are more likely to lead to interesting results. However, domain or SUT specific distances cannot be assumed to be available in the general case. The development, implementation, and validation of such distance metrics is likely to be costly in terms of time and resources. Moreover, particular assumptions that developers have about the behaviour of their systems, and assumptions about their application domain, may not be accurate. 

Note that the method will find the boundary regardless of the distance metric used. The distance that is used will impact the way the input space is understood and explored, and will therefore have an impact on the outcome. But the method will find a boundary regardless of the distance metric used. 
In light of the potential trade-offs, our validation will investigate four distances, as mentioned above: a domain-agnostic distance metric (NCD), a domain-agnostic but representation specific distance metric (Levensthein), and two domain-specific distances (Most Significant Integer Distance and Approximate Day Distance). 

% subsection distances (end)

% section validation (end)

\section{Results and Analysis} % (fold)
\label{sec:results}

\subsection{Initial Analysis} % (fold)
\label{sub:narrow}

The method was initially applied to generate input data for the DateParser~\footnote{https://github.com/invenia/DateParser.jl} module in the Julia programming language. 

The first step consisted of the development of a date generator in G\"{o}delTest. The generation process was optimised using NMCS. The fitness function for the search was the minimum distance between a new datum being generated and the existing data set, and the fitness function was to be maximised. At each point, the NMCS algorithm evaluated only two choices: thus the settings were selected to focus on generation time rather than accuracy improvements. The result was a set of $10$ valid dates, that are assumed to be at different points along the valid-invalid boundary. An illustration of the first step can be seen in Figure~\ref{fig:step1}. For step 1, all four distances described above were used, resulting in four initial valid data sets.

The second step consisted of mutating each of the $10$ valid dates in each of the initial sets. The result was a pair of valid invalid sets for each initial point, that describe the boundary in the neighbourhood of that point. The mutations were applied on the genotype, i.e.\ on the string representation of each data point. The mutations applied were:
\begin{itemize}
    \item \textbf{IncreaseInt}: The mutation operator finds an integer figure, i.e.\ between $0$ and $9$, in the string and increases its value by one. If the result is not a figure, it will be removed from the string, i.e.\ the mutation operator may result in a change in string size. 
    \item \textbf{DecreaseInt}: The mutation operator finds a integer figure, i.e.\ between $0$ and $9$, in the string and decreases its value by one.If the result is not a figure, it will be removed from the string, i.e.\ the mutation operator may result in a change in string size. 
    \item \textbf{IncreaseInKeepingSize}: The mutation operator finds a integer figure, i.e.\ between $0$ and $9$, in the string and increases its value by one. The result is always subjected to a modulo $10$ operation. As a result, the mutation operator will maintain the size of the mutated string.
    \item \textbf{DecreaseInKeepingSize}: The mutation operator finds a integer figure, i.e.\ between $0$ and $9$, in the string and decreases its value by one. The result is always subjected to a modulo $10$ operation. As a result, the mutation operator will maintain the size of the mutated string.
\end{itemize}

Below are the variables that describe each execution of the current method:
\begin{itemize}
    \item Initial Test set: The starting set of test cases for the current SUT.
    \item Distance: The type of distance use in the fitness function for data generation.
    \item Tset: The set of valid test cases produced by the generator and the starting point for mutations. 
    \item Mutation Operators: The set of mutation operators applied  to each element of the Tset.
    \item Valid-Invalid sets: the paired sets of valid and invalid mutated test cases associated with each element of the Tset. 
    \item Random set: A set of randomly generated valid test cases used for assessment.
    \item Generated Invalid set: A set of automatically generated invalid test cases. Given that the SUT in this case was a date generator, a set of invalid dates that are close to the boundary could be generated for analysis purposes. 
\end{itemize}

We computed the minimum distances between the paired valid-invalid sets, as well as the distance between the valid set and the initial set, the valid set and the random set, and the valid set and the generated invalid set. 

If the distance between the valid set and its invalid pair is greater than the distance between the valid set and any of the other sets, we conclude that there are test cases between those sets and the boundary. 

If the distance between the paired valid-invalid sets is small that the distances between the valid set and all the other sets, we conclude that no test cases were found between those sets and the boundary itself. In this case, we can conclude that the paired sets describe the boundary between the valid and invalid regions of the space accurately, for the given region of the boundary. 

\begin{figure*}
	\centering
		\includegraphics[scale=0.26]{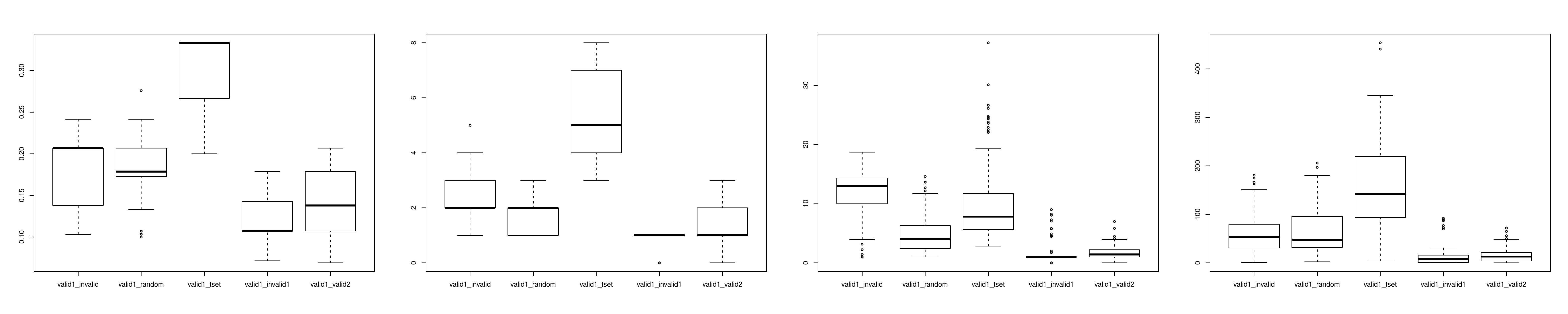}
	\caption{Overview of the distances between various sets. The valid set was developed using G\"{o}delTest, maximising the minimum distance between new candidates and the valid set. The distance used was NCD. The mutation operators were IncreaseInt and DecreaseInt.}
	\label{fig:ncd1}
\end{figure*}

\begin{figure*}
	\centering
		\includegraphics[scale=0.26]{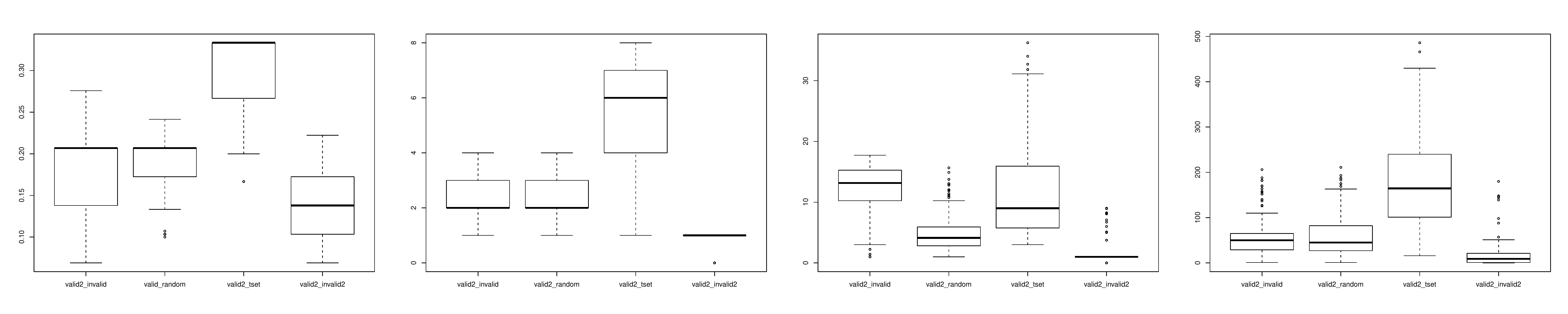}
	\caption{Overview of the distances between various sets. The valid set was developed using G\"{o}delTest, maximising the minimum distance between new candidates and the valid set. The distance used was normalised compression distance. The mutation operators were IncreaseIntKeepingSize and DecreaseIntKeepingSize.}
	\label{fig:ncd2}
\end{figure*}

The results in Figures~\ref{fig:ncd1} and~\ref{fig:ncd2} show examples of the results for the paired valid-invalid test sets obtained for the DateParser package, using NCD as a distance for data generation. Each figure shows four graphs, corresponding to the data being analysed using each of the four distance metrics. The data in Figure~\ref{fig:ncd1} was obtained using the IncreaseInt and DecreaseInt mutation operators, while the data in Figure~\ref{fig:ncd2} was obtained using the IncreaseIntKeepingSize and DecreaseIntKeepingSize mutation operators. The four diagrams in each figure are, in order from left to right: valid1-invalid (the comparison between the Valid set and the Generated Invalid set), valid1-random (the comparison between the Valid set and the Random set), valid1-Tset (the comparison between the Valid set and the Tset), valid1-invalid1 (the comparison between the valid and invalid paired sets), and valid1-valid2 (the comparison between two Valid sets obtained with different distances). The results are consistent with the other applications. The distance between the valid and invalid sets, seen in the graphs as valid-invalid, is consistently smaller than the distance between the valid sets and the other sets under comparison, according to all the distances used for the assessment. 
%bmr: I will redo those figure to be clearer. 

\begin{figure}
	\centering
		\includegraphics[scale=0.39]{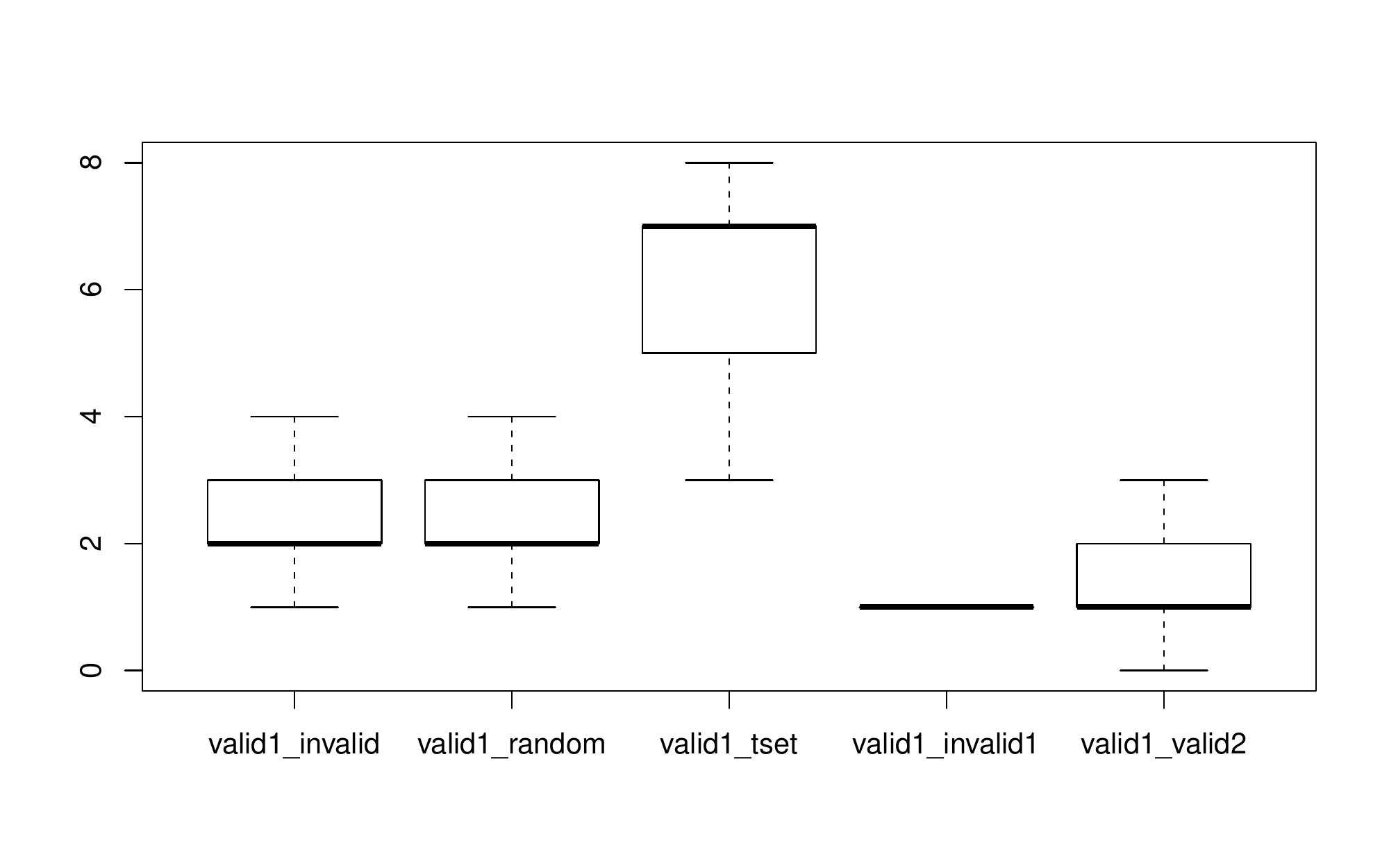}
	\caption{Detailed view of some of the results. The valid set was developed using NCD, while the analysis was conducted using Levenshtein distance. The mutation operators were IncreaseInt and DecreaseInt.}
	\label{fig:ncd_lev1}
\end{figure}

Figure~\ref{fig:ncd_lev1} shows a detailed view of some of the results. The results in the figure were obtained by generating the initial data set from G\"{o}delTest using NCD as a distance metric, and using IncreaseInt and DecreaseInt for mutation operators. The data was analysed using the Levenshtein distance metric. The box-plots show, the comparison between the Mutated Valid Set (MVS) and the reference invalid set (invalid), the random valid set (random), the initial test set (Tset), the Mutated Invalid Set (MIS, seen in the figure as invalid1), and the MVS obtained using alternative mutation operators (in the graph as valid2), respectively. 

The results show that the mutated valid set is closer to the mutated invalid set than to any of the other test case sets, according to the distance metric used. Since the MVS and MIS only contain valid and invalid test cases, respectively, we conclude that the boundary between the valid and invalid regions of the space, in the given neighbourhood, lies between those two sets.  As a result, we can conclude that the paired valid-invalid test sets, the MVS and the MIS, approximate the boundary better than any of the alternatives currently considered. 

The box plot on the far right of Figure~\ref{fig:ncd_lev1} shows a comparison between the MVS obtained with the current set of mutation operators, and the MVS obtained with the alternative set of mutation operators. This serves as a sanity check: since the mutation operators act on the data set in similar ways and operate under similar restriction, it would be expected that the resulting mutated sets would also not be very different. This serves to reinforce the conclusion that the results seen are a result of the method being employed, rather than being an artefact of the non-deterministic data generation.  

The results presented here indicate that the method proposed was able to find a valid-invalid pair of test sets that describes the boundary between the valid and invalid regions of the space. This conclusion applies regardless of the distances being used for data generation and for data analysis. While differences between the distances exist, the overall conclusions stand. 

% subsection narrow (end)

\subsection{Validation in Practice} % (fold)
\label{sub:validation_practice}

In practice, validation consisted in measuring the distances between the mutated valid set (MVS), i.e.\ the valid pair of the mutated sets, and several sets of test candidates. These sets are presented below:
\begin{itemize}
    \item \textbf{Reference Invalid Set (invalid)}: The reference invalid set contains a test cases that are known to be invalid, but close to the valid boundary. For the DateParser, this set we generated this set based on domain knowledge, for the purpose of analysis. For a generic SUT, however, such a set will not be available, and follow-up validations will not rely on this data. 
    \item \textbf{Random Valid Set (random)}: The random valid set contains randomly generated valid test cases. The purpose of this set is purely for analysis. Comparing the MVS against a randomly generated test set allows a certain degree of confidence that the effects seen are a result of the optimisation process, and not an artefact of the non-deterministic data generation. 
    \item \textbf{Initial test set (Tset)}: This set contains the initial tests associated with a particular SUT. The comparison between the MVS and the Tset shows the distance between the starting point of the search for the boundary and the final state.
    \item \textbf{Mutated Invalid Set}: The mutated invalid set (MIS) is the pair of the Mutated Valid Set, resulting from the property switching search. We expect that the distance between the MVS and the MIS is significantly smaller than the distance between the MVS and other sets, since the MVS and MIS are the result of step-wise mutations around the same region of the space. 
    \item \textbf{Mutated Valid Set alternative (valid2)}: For the DateParser, we used two sets of mutation operators to obtain the MVS and MIS. This is a sanity check comparison, to assess if the two sets of mutation operators result in very different results. The mutation operators used are discussed above, and fulfil the same roles with small differences. 
\end{itemize}

Note that not all these sets are available for the wider validation. For example, developing a set of reference invalid candidates, that are close to the valid boundary, was a relatively simple task for the DateParser package. For a generic SUT, however, we cannot assume that such a reference set is available or can be developed with a reasonable degree of effort or confidence. As a result, the wider validation efforts do not assume that such a set is available and do not include it in the analysis.

%subsection validation_practice (end)

\subsection{Widening the analysis} % (fold)
\label{sub:wide}

For the date generator, quite a lot of domain-specific understanding was available, allowing the development of specialised distance function and the generation of invalid, but reasonable, data for analysis. 

In a generic SUT, such domain-specific information cannot be assumed to be available. As a result, we widened the analysis, to include three more SUTs from the Julia package library. The SUTs included in the widened analysis are described below:

\begin{itemize}
    \item \textbf{Regex:} This module allows users to generate regular expressions, for use in other Julia applications, from a given input string. If the given input string can be parsed by Regex, the input is considered valid. 
    \item \textbf{XMLDict:} The module allows users to parse an xml string and returns an equivalent Dict structure in Julia. If the given input string can be parsed into a valid Dict structure, the input is considered valid. For this SUT, a data generator was developed that first created a valid Dict structure in the Julia language and then transformed it into the equivalent xml representation.
    \item \textbf{JSON:} The module allows users to convert Julia structures to the JSON format for storage and transmission to other subsystems. For this evaluation, the JSON string was considered the input for the JSON package. If the JSON package was able to parse the input string and produce a valid Dict structure, the input string was considered valid. The same Dict structure generator was employed to create valid Dict structures and transform them into the equivalent JSON representation.
\end{itemize}

Note that, for all the SUTs above, the semantics of the inputs are not discussed. The only assessment of validity consists of the ability or inability of the package in question to parse the input string. Note also that, since the Dict and Regex objects are less structured that the date in the previous system, the analysis was performed only using general distances, i.e.\ the domain-agnostic normalised compression distance (NCD), and the representation specific Levenshtein distance. Since all the input values were strings, a string distance is a reasonable choice. 

For each of the SUTs, step 1 consisted of the development of a data generator that produced valid data of the respective type, i.e.\ Regex or Dict, and then converted it to a valid string according to the respective representation. The data generation was optimised as in the previous case: NMCS was used. The fitness function was the maximised minimum distance between the generated candidate and the existing set of valid test cases. The NMCS only assessed two choices at each choice point, and was optimised for quick, rather than accurate, data generation.

The mutation operators used for all the data were string-specific:
\begin{itemize}
    \item \textbf{DeleteChars(1):} One random character in the string was deleted. 
    \item \textbf{CopyChars(1):} One random character in the string was duplicated. 
\end{itemize}

Below are the variables that describe each execution of the current method:
\begin{itemize}
    \item SUT: the Julia package that forms the system under test.
    \item Initial Test set: The starting set of test cases for the current SUT. For each of the SUTs discussed here, we randomly generated one valid test case to form the initial test set. 
    \item Distance: The type of distance use in the fitness function for data generation. For the SUTs here, only NCD and Levenshtein distances were used for both data generation and analysis. 
    \item Tset: The set of valid test cases produced by the generator and the starting point for mutations. 
    \item Mutation Operators: The set of mutation operators applied  to each element of the Tset.
    \item Valid-Invalid sets: the paired sets of valid and invalid mutated test cases associated with each element of the Tset. 
    \item Random set: A set of randomly generated valid test cases used for assessment.
\end{itemize}

\begin{figure}
	\centering
		\includegraphics[scale=0.13]{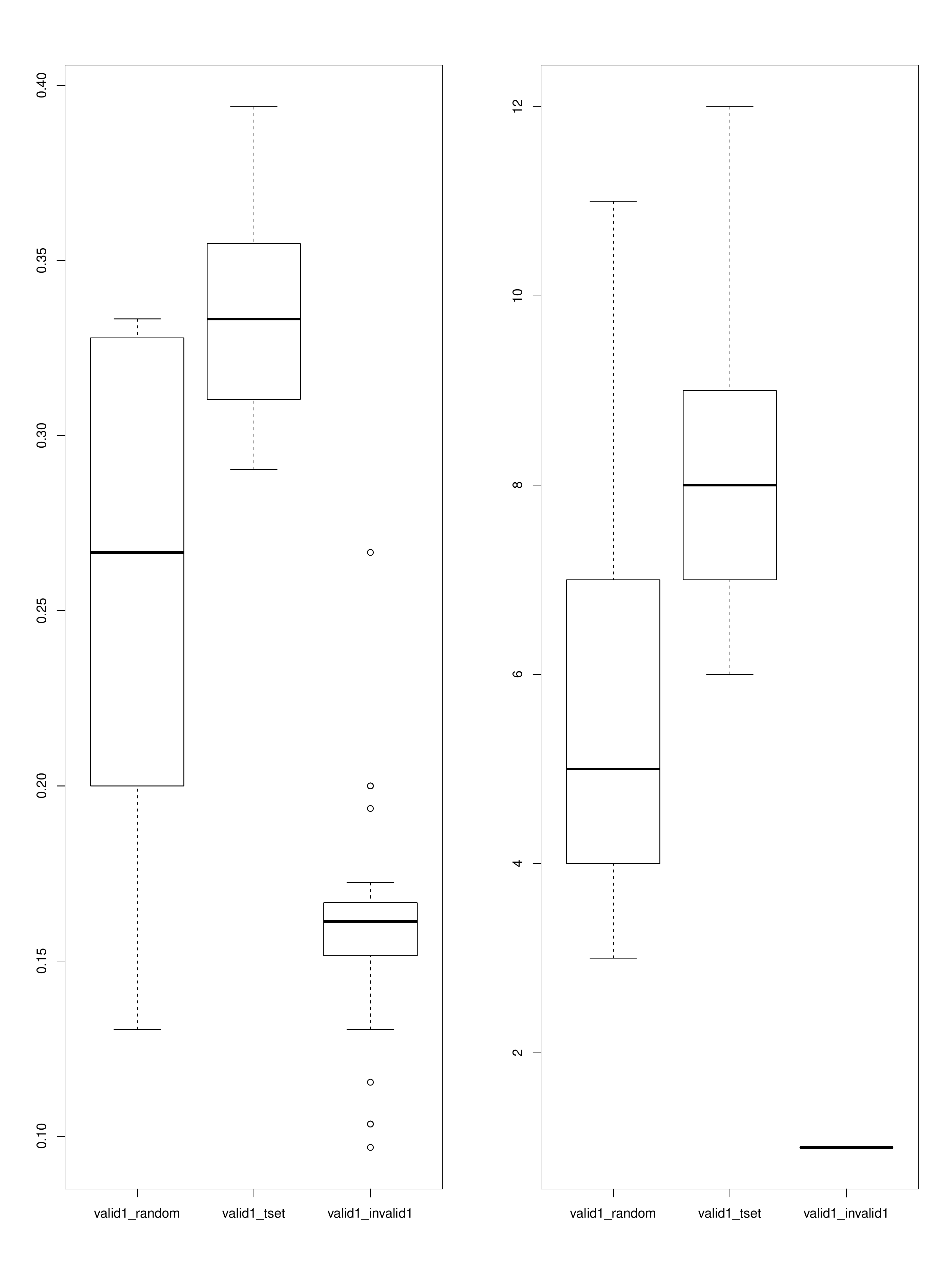}
		\includegraphics[scale=0.13]{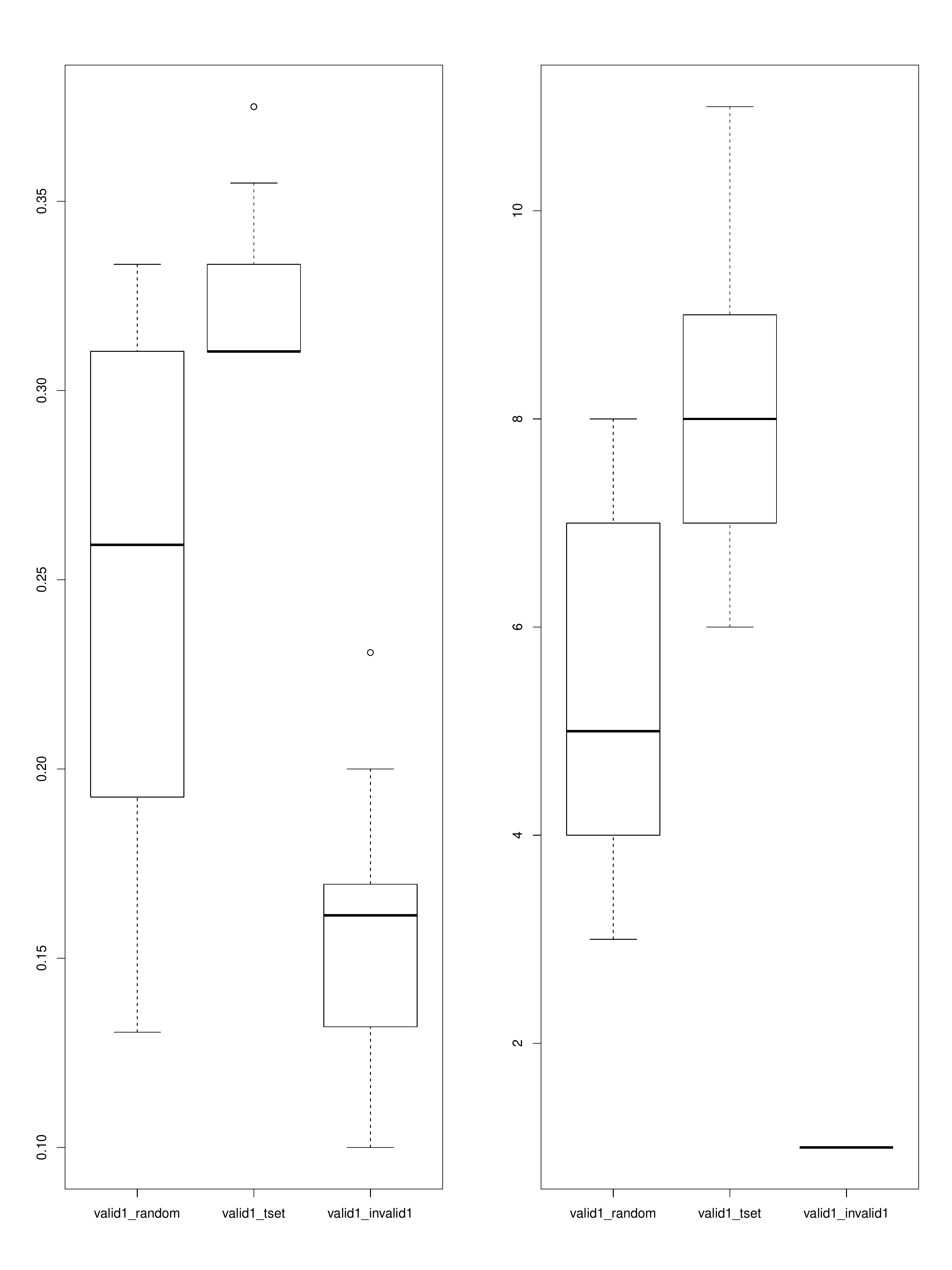}
	\caption{Overview of the distances between various sets for the Regex package. The valid set was developed using G\"{o}delTest, maximising the minimum distance between new candidates and the valid set. The distance used was NCD. For each diagram, the box-plots are, respectively: On the left, the distance between the valid test and the random set; in the middle, the distance between the valid set and the Tset; on the right, the distance between the valid set and the invalid set.}
	\label{fig:regex}
\end{figure}

\begin{figure}
	\centering
		\includegraphics[scale=0.13]{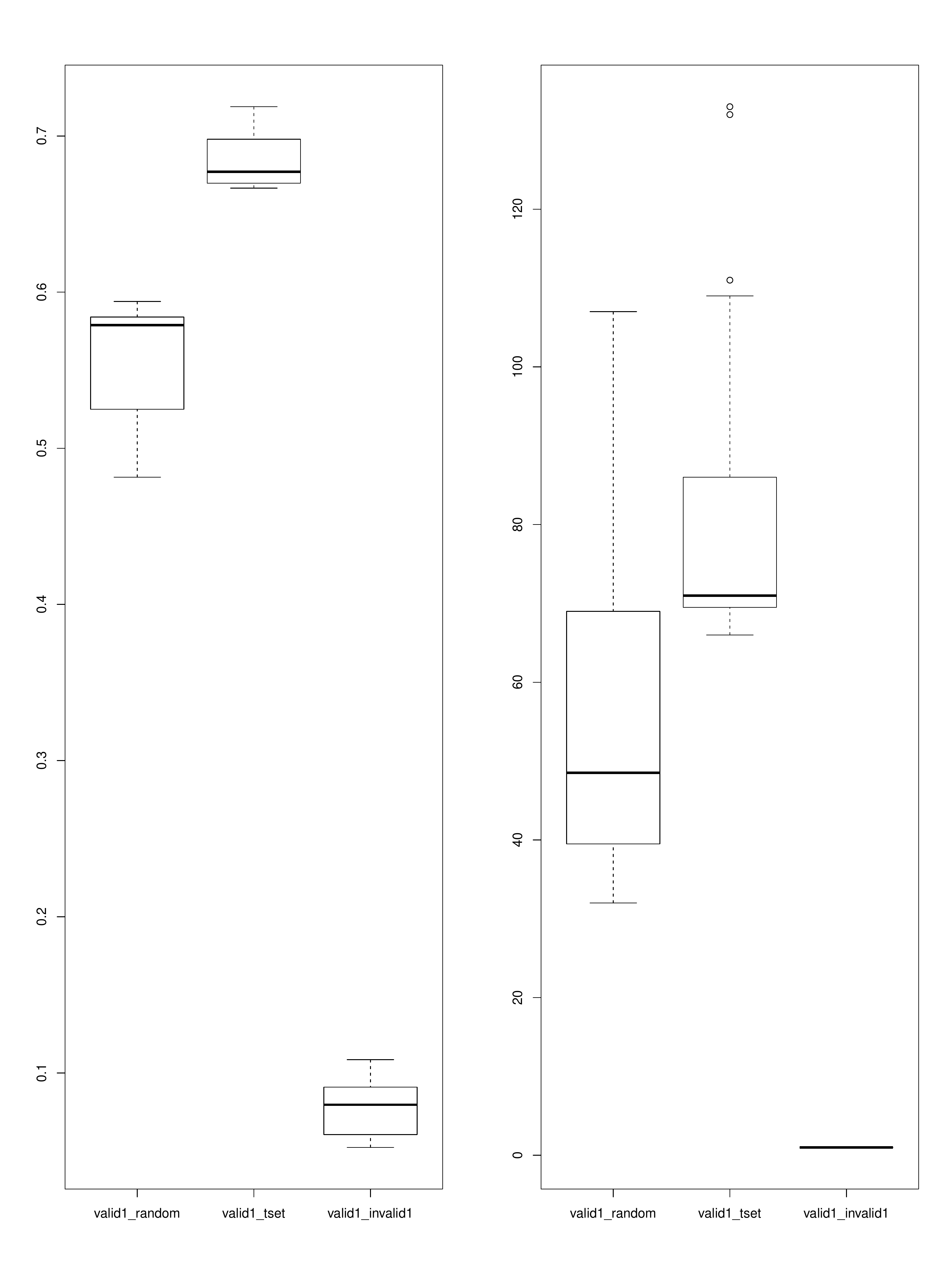}
		\includegraphics[scale=0.13]{img/xmldict_ncd.pdf}
	\caption{Overview of the distances between various sets for the XMLDict package. The valid set was developed using G\"{o}delTest, maximising the minimum distance between new candidates and the valid set. The distance used was NCD. For each diagram, the box-plots are, respectively: On the left, the distance between the valid test and the random set; in the middle, the distance between the valid set and the Tset; on the right, the distance between the valid set and the invalid set.}
	\label{fig:xmldict}
\end{figure}

\begin{figure}
	\centering
		\includegraphics[scale=0.13]{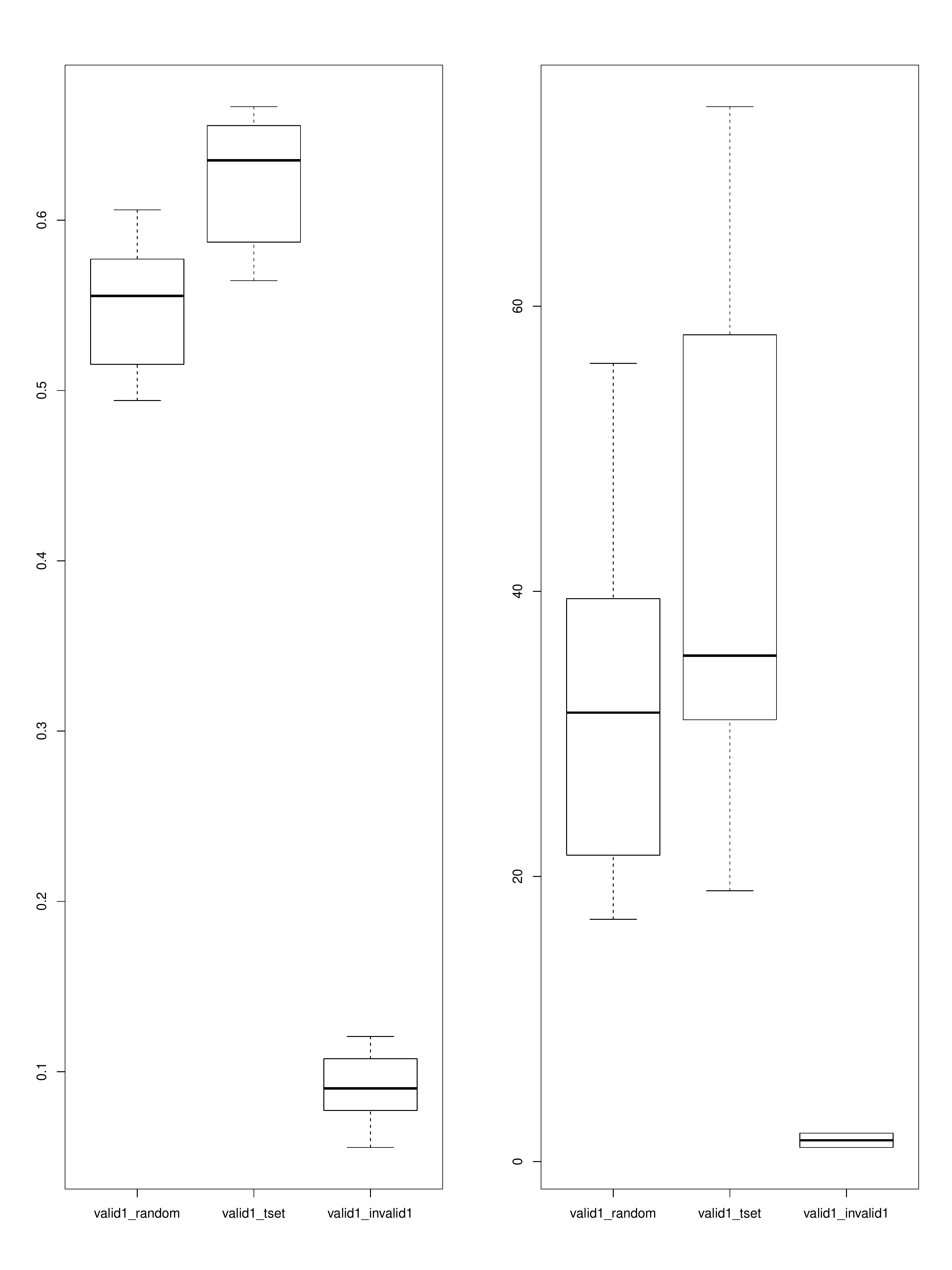}
		\includegraphics[scale=0.13]{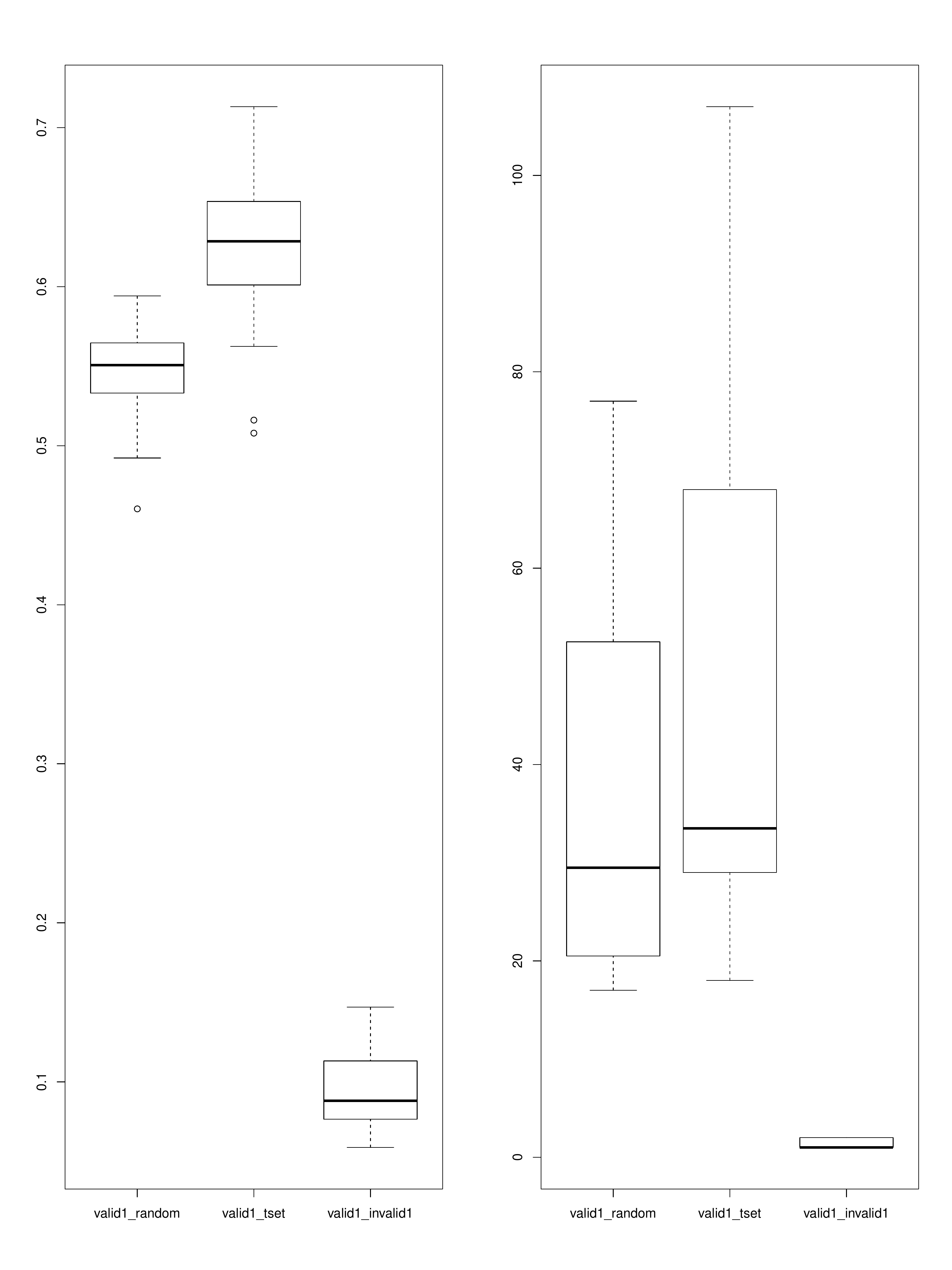}
	\caption{Overview of the distances between various sets for the JSON package. The valid set was developed using G\"{o}delTest, maximising the minimum distance between new candidates and the valid set. The distance used was NCD. For each diagram, the box-plots are, respectively: On the left, the distance between the valid test and the random set; in the middle, the distance between the valid set and the Tset; on the right, the distance between the valid set and the invalid set.}
	\label{fig:json}
\end{figure}

Figures~\ref{fig:regex},~\ref{fig:xmldict}, and~\ref{fig:json} show the results of the method being applied on the three SUTs described above in detail. For these SUTs, domain-specific distance metrics were not used, either for data generation or for analysis. The two sub-figures on the left of each figure correspond to the data generated using NCD and analysed using NCS and the Levensthein distance respectively. The two sub-figures on the right correspond to data generated using the Levensthein distance and analysed using NCS and Levensthein distance respectively. For each diagram, the box-plots show the comparison between the Mutated Valid Set (MVS) and the randomly generated valid set, the initial test set (Tset), and the Mutated Invalid Set (MIS), respectively. 

We observe consistently, that regardless of the distance metrics being used for data generation and analysis, and regardless of the SUT, the distance between the paired valid-invalid sets, the MVS and the MIS, is consistently and significantly smaller than the other distances. Since the paired valid-invalid sets only consist of valid and invalid test cases respectively, we conclude that the boundary between the valid and invalid regions of the space, for the given neighbourhood, lies between the test cases in those two sets. 

%In all cases, the rightmost plot, describing the distance between the paired valid-invalid sets is significantly smaller than the distance between the valid and the random or initial sets. 
Thus, we conclude that, even in the absence of domain specific distances, it is possible to find the boundary between the valid and invalid sets and describe it for a given region. 

% subsection wide (end)

% section results (end)

\section{Discussion and Future Work} % (fold)
\label{sec:discussion}

In this paper, we have described a way of determining where the boundary between the valid and invalid regions of the input space lies. This boundary is described by a pair of test sets, one containing valid test cases and the other containing invalid test cases. Since the distance between these sets is minimal, we conclude that the pair describes the boundary. 

We used a set of distance metrics, ranging from the general to the domain specific. For the cases we studied in the current paper, the exact distance metric used appeared to have some effect, but not enough to change the conclusions of the work. However, we did notice variations in the average distance between the paired valid-invalid test sets. This seems to suggest that, while the method proposed works regardless of the distance metric used, some distance metrics may result in a better approximation than others. Note, however, that domain-specific distances may require additional development and may not always be available. 

Conversely, Normalised Compression Distance (NCD) is a generic type of distance that can be applied on any type of data. It can be argued, however, that NCD is computationally demanding, especially in situations where a large number of comparisons is required. For example, when computing the minimum distances between two sets, as proposed in this paper, the number of evaluations increases dramatically with the size of the sets. Computationally expensive distance metrics may be impractical for large data sets. 

Levenshtein distance seems to perform well in the SUTs we have considered for this work. However, this is applicable for SUTs where the string encoding of the input appears to have some semantic significance. It is difficult to claim that this approach is generalizable beyond such cases, and it may not be trivial to determine if this distance is significant or not for generic SUTs.

Thus, a set of trade-offs need to be made between the time and effort required to develop specific measurements, the quality and quantity of the resulting test sets, and the computational resources that can be made available for such methods.

All the distances we used in the paper measure the similarity of the genotypes of different solutions, and all the mutations referred to mutating the representations, in particular the string representation, of those solutions. Distance metrics defined on the common representations allow a more generic approach to test data generation. However, test cases can be developed that are close in terms of genotype, but very distant in terms of phenotype. The exact impact of the use of genotype or phenotype focused distances is difficult to assess, but could lead to interesting results in future work.

Note also that, for this work, we used sets of relatively simple mutation operators. The development of domain or application specific mutation operators is also possible and could be beneficial in terms of computational resources and the time needed to explore the boundary. More in-depth research is needed to investigate this possibility, and to fully explore the impact of mutation operators on the property-switching search phase.

The method presented in this paper focuses on a number of discrete points that are close to the boundary: the test cases contained in the Tset. The search for the boundary starts from those discrete points and investigates the boundary between the valid and invalid regions of the space in the neighbourhood of those points. At the moment, we have no way of evaluating how much of the boundary is covered, and implicitly what regions of the boundary are still not covered, or how much of the boundary is still unknown. Thus, we are able to find the boundary, but have no way of describing the entirety of the boundary at present.

Property-switching search could be used, augmented with some means of directional guidance, to move along the boundary and connect the paired valid-invalid test sets in different neighbourhoods. Further research would be needed to described the boundary in its entirety and determine if there are sections that are not currently covered. This is especially problematic for SUTs that have non-contiguous valid spaces, and where simply tracing along the boundary does not guarantee that the entire boundary will be covered. The currently proposed method reduces the likelihood of missing large pockets of the valid space, by generating several starting point. Nevertheless, the possibility exists that unlikely, but still valid, test cases in non-contiguous regions of the space might be missed.

% section discussion (end)

\section{Conclusions}
\label{sec:conclusions}

This paper proposes a method for finding and describing the boundary between the valid and invalid regions of an input space. We start from automated data generation, using G\"{o}delTest to generate valid test cases. Nested Monte-Carlo Search, maximising the minimum distance between new data and existing data sets, is used to push data being generated towards the boundaries of the valid region of the space. 

The valid test cases are then mutated, using property-switching search, to generate paired valid-invalid test sets. Since we have a pair of valid and invalid test sets, we can assume that the boundary between the valid and invalid regions of the space lies between those two sets. And since the distance between those sets is relatively small, we can assume that the paired valid-invalid test sets can be used to describe that boundary.

%\section{References}

\bibliographystyle{elsarticle-num}
\bibliography{p8}

\end{document}